\documentclass{article}
\usepackage{hq2kpro}
\newcommand{\kpnn}  {$K^+ \! \rightarrow \! \pi^+ \nu \overline{\nu}$ }
\newcommand{\kmng}  {$K^+ \! \rightarrow \! \mu^+ \nu_\mu \gamma$ }
\newcommand{\kppg}  {$K^+ \! \rightarrow \! \pi^+ \pi^0 \gamma$ }
\newcommand{\kplus} {$K^+ \! $}
\newcommand{\vtd}   {$V_{td} \! $}
\newcommand{\kpen}  {$K^+ \! \rightarrow \! \pi^0 e^+ \nu_e$ }
\newcommand{\kpmn}  {$K^+ \! \rightarrow \! \pi^0 \mu^+ \nu_{\mu}$ }

\begin{document}
\title{ 
Rare $K^+$ Decays from Experiment E787
}
\author{
Vivek~Jain        \\
{\em Brookhaven National Laboratory} \\
}
\maketitle
\baselineskip=11.6pt
\begin{abstract}

   This paper presents the latest results from experiment E787, at
Brookhaven National Laboratory, on \kpnn and radiative \kplus \space
decays. The result for \kpnn uses data collected in runs taken during
1995, 1996 and 1997.  In addition, we discuss plans for future
measurements of \kpnn.

\end{abstract}
\baselineskip=14pt
\section{Theoretical Background}
\subsection{\kpnn}

This decay mode is a very clean probe of the dynamics of the
Cabbibo-Kobayashi-Maskawa (CKM) mixing matrix. It is a $\Delta S=1$
process which proceeds via the box diagram and electroweak penguin
diagrams\cite{buras}. Since the top quark is so much more massive than
the other up-type quarks, its contribution dominates the decay rate.
In addition, long distance effects are small and the hadronic matrix
element can be obtained from the semi-leptonic decay \kpen
($K_{e3}$). As a result, the theoretical uncertainty in understanding
the decay rate is rather low (theoretical error is $\approx 7\%$, most
of it is due to the charm quark contribution). The rate can be
expressed (in the Standard Model) as,

\begin{equation}
B(K^+ \! \rightarrow \! \pi^+ \nu \overline{\nu}) = 
\frac{\kappa_+ \alpha^2 B(K_{e3})} {2 \pi^2 \sin^4 \theta_W |V_{us}|^2}
      \Sigma_l  |X_t\lambda_t + X_c^l\lambda_c|^2
\label{formula} 
\end{equation}

\noindent where, $\kappa_+$ is the isospin correction, $\lambda_t =
V_{ts}V^*_{td}$, $\lambda_c = V_{cs}V^*_{cd}$, and $X_{t(c)}$ are
Inami-Lim functions\cite{buras}, and the sum $\Sigma_l$ is over the
three neutrino flavours.  Using measurements from the K and B systems, to
determine parameters of the CKM matrix, which are used as inputs, one
can predict, \\

$0.5 \times 10^{-10} < $ \space B(\kpnn) \space $  < 1.2 \times 10^{-10}$ \\

This range is due to the uncertainties in the input parameters.  Since
the top quark dominates the proceedings, this decay mode provides a
very clean measurement of \vtd.  This measurement is complementary to
the measurement of \vtd \space from $B^0_{d} \overline B^0_{d}$
mixing. A difference in the value of \vtd \space extracted from these
two sources, could suggest new physics, since any new phenomena would,
in general, affect the K and B systems differently.

\subsection{Radiative \kplus \space decays}

Radiative \kplus \space decays come from two sources, (a) the
``pedestrian'' Inner \\ Bremsstrahlung decay, where a charged decay
daughter emits a photon, and (b) the more interesting structure
dependent (SD) part. In this paper, we are concerned with the SD
contribution.

The latter source is a good testing ground for Chiral Perturbation
theory ($\chi$PT), and is also important for measuring long distance
contributions to other decays of interest, {\it e.g.},
$K^0_L~\rightarrow~l^+l^-\gamma\gamma$ is a background to
$K^0_L~\rightarrow~\pi^0l^+l^-$.

For \kmng, form factors for the SD part, ${\rm F_A}$ and ${\rm F_V}$
are predicted by $\chi$PT. \kppg is interesting in its own right;
decay rates for \kplus, $K_L$ and $K_S$ could be similar, even though
the decay rates for the non-photonic final state ($\pi\pi$) are
very dissimilar, also, some SUSY models predict that $K^+$ and $K^-$
decay rates are unequal, leading to direct CP violation.
\section{Search for \kpnn}

\subsection{Experimental Design}

     Since the expected branching fraction is $\approx 10^{-10}$,
backgrounds are {\bf the} major concern, consequently, the entire
experiment and the analysis techniques are geared towards reducing
backgrounds while maintaining a reasonable detection efficiency.

\begin{figure}[ht]
\vspace{5.5cm}
\includegraphics{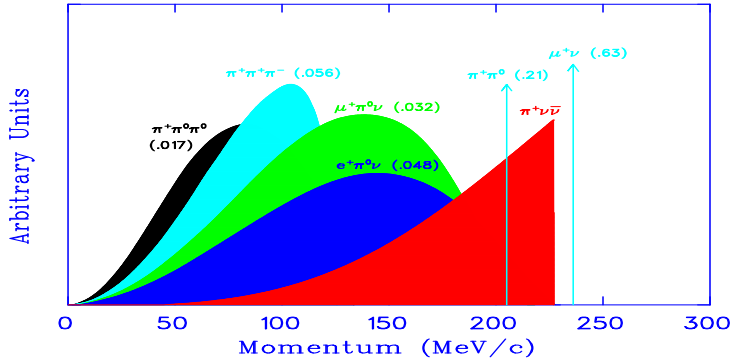}
\caption {\it Momentum spectrum of charged tracks produced in \kplus \space decays 
    \label{phase}}
\end{figure}

    The two major backgrounds are $K^+ \rightarrow \mu^+ \nu$
(BR=64\%) and $K^+ \rightarrow \pi^+ \pi^0$ (BR=21\%). Other sources,
{\it e.g.}, \kpen, \kpmn, etc., are not as daunting. Fig.~\ref{phase}
shows the momentum spectrum of charged tracks in major \kplus \space
decays.  The experimental strategy is set by the characteristics of
the backgrounds.

Since the two largest decay modes give off mono-energetic charged
daughters ($P_{\pi} = 205$ MeV/c and $P_{\mu} = 236$ MeV/c,
respectively), we stop the incident beam and let the \kplus \space decay at
rest. If we maintain very good kinematic resolution, and search in a
momentum region inside this range, we have one discriminant against
these two background sources (the momentum spectrum in
Fig.~\ref{phase} for the $\pi^+$ in \kpnn decays has been enhanced by
$\sim 10^{10}$). We also use the energy and range of the tracks as
part of the kinematic discriminant.

   To further discriminate against muons, we detect the decay chain, \\
 $\pi^+ \! \rightarrow\ \! \mu^+ \!\rightarrow \! e^+$. To do this, we require
that the outgoing $\pi^+/\mu^+$ stop in the detector and decay. The 
$\pi^+$ will generate a $\mu^+$ and an $e^+$, whereas a muon will only
give rise to an $e^+$.

   Most \kplus \space decays produce a $\pi^0$ in the final state, so
we designed the experiment to have very good photon veto capabilities,
e.g., the average rejection for the $\pi^0$ in $K^+ \rightarrow \pi^+
\pi^0$ is about $10^6$.

   There are other sources, {\it e.g.} pions in the incident beam
which scatter in the target and fall inside the signal region, charge
exchange where a \kplus \space turns into a $K_L$, and the latter
decays semi-leptonically. Such sources can be cut out by requiring
that the outgoing pion be detected a few nano-seconds after the
incident \kplus \space enters the target. This not only drastically
reduces the possibility that an incoming pion scatters into the
fiducial region, but also ensures that the incoming \kplus \space
decays at rest.

   The E787 has been described in detail elsewhere\cite{detector}. Its
main features are a {\it degrader} to slow the incident \kplus's, so
that they come to rest in an {\it active target} which is mainly
composed of 413 5mm scintillating fibers. The target is surrounded by
a {\it drift chamber}, and a {\it range stack} composed of 21 layers
of plastic scintillators. The outgoing daughter pions are required to
stop in the range stack, where we detect the $\pi \rightarrow \mu
\rightarrow e$ decay chain using {\it 500 MHz transient
digitizers}. Photons are detected with an extensive system of {\it
photon veto detectors}, e.g., Pb-Scintillator calorimeter in the
central region, CsI crystals in the endcap regions, etc. Track
momentum is measured in the presence of a 1T solenoidal magnetic
field.

\subsection{``Online'' improvements} \label{section:dataset}

Since we first published the observation of this decay
mode\cite{first} using data collected in 1995, we have taken more data
during runs in 1996, 1997 and 1998.  We have made many improvements;
lowering the momentum of the \kplus \space beam, which results in a
higher fraction of incident \kplus \space stopping in the target and a
reduction in background hits from \kplus \space interactions in the
degrader, improvements to the trigger, increased acceptance for
detecting $\mu \rightarrow e$ decays, etc.

In Table~\ref{data}, we present some details of the data runs in 1995-97.

\begin{table}[t]
\centering
\caption{ \it 1995-97 dataset
}
\vskip 0.1 in
\begin{tabular}{|l|c|c|c|c|} \hline
          & 1995 & 1996 & 1997 & Total \\
\hline
\hline
Length of run (weeks) & 25 & 17 & 8  &  \\
Triggers ($10^{12}$)  & $\sim 1.53$ & $\sim 1.16$ & $\sim 0.59$ & $\sim 3.28$ \\
No. of evts. to tape  &              &              & & $\sim 3.1\times 10^8$ \\ 
\hline
\end{tabular}
\label{data}
\end{table}

Even though the data runs were shorter in 1996 and 1997, we still
collected more data as compared to the run in 1995. This was entirely
due to the improvements outlined above.

\section{Offline Analysis}

    Since the backgrounds are orders of magnitude larger than the
signal, the analysis has to be designed very carefully. Some of the
main features are, 

(a) {\it Blind Analysis} The signal region is hidden (by inverting
cuts) while cuts are developed and background levels are estimated.
To avoid any bias in the background estimates, we develop cuts on 1/3
of the dataset and take the (actual) background level from the
remaining 2/3 of the data and extrapolate the latter to the full
sensitivity.

(b) {\it ``Bifurcated'' Analysis} In the analysis, we make an {\it ``a
priori''} identification of the background sources and then develop
at least two independent cuts for each source. In this manner, we can
measure the rejection of each cut using the data itself. For instance,
$K^+ \rightarrow \pi^+ \pi^0$ is rejected using kinematic cuts based
on measuring the Range, Momentum and Energy of the outgoing $\pi^+$
AND photon veto cuts which work on detecting the $\pi^0$. Similarly,
$K^+ \rightarrow \mu^+ \nu_\mu$ is rejected by applying kinematic cuts
on the $\mu$ and by requiring the presence of the $\pi^+ \rightarrow \mu^+
\rightarrow e^+ $ decay chain.

    The goals for the analysis of the 1996-97 datasets and the
re-analysis of the 1995 dataset were (a) to increase rejection so as
to keep the total background at the same level as in the original
result\cite{first}, (b) maintain or (possibly) increase detection
efficiency, and (c) devise methods to increase background samples
which would lead to a better understanding and a more precise
estimation of the background.

  A lot of improvements were made in the analysis; tracking changes in
the target, range stack and drift chamber improved the resolution in
range and momentum, a better electron (from $\mu$ decay) finding
algorithm was put in place, new cuts were devised to reject \kplus
\space decays in flight, etc. For instance, the new analysis had a
30\% larger rejection (with the same acceptance) against muon
backgrounds.

The background estimates for the 1995-97 dataset are presented in
Table\ref{bkgd}.

\begin{table}[ht]
\centering
\caption{ \it Background Estimate for 1995-97 datasets}
\vskip 0.1 in
\begin{tabular}{|l|c|}
\hline\hline
                    & Total \\
\hline
$K^+ \rightarrow \pi^+ \pi^0$    & 0.020 $\pm$ 0.01 \\
$K^+ \rightarrow \mu^+ \nu_\mu$ & 0.030 $\pm$ 0.01 \\
1(\&2)-beam                     & 0.020 $\pm$ 0.02 \\
Charge Exchange                 & 0.010 $\pm$ 0.01 \\
\hline
Total                           & 0.08 $\pm$ 0.02 \\ 
\hline \hline
\end{tabular}
\label{bkgd}
\end{table}

The background estimate in the published result\cite{first} was
$0.08\pm 0.03$, which implies that we increased rejection by a factor
of about 2.2. At the same time, we were able to increase acceptance
from 0.16\% to 0.21\% (an increase of almost 30\%).

\begin{figure}[ht] 
\vspace{8.0cm}
\includegraphics{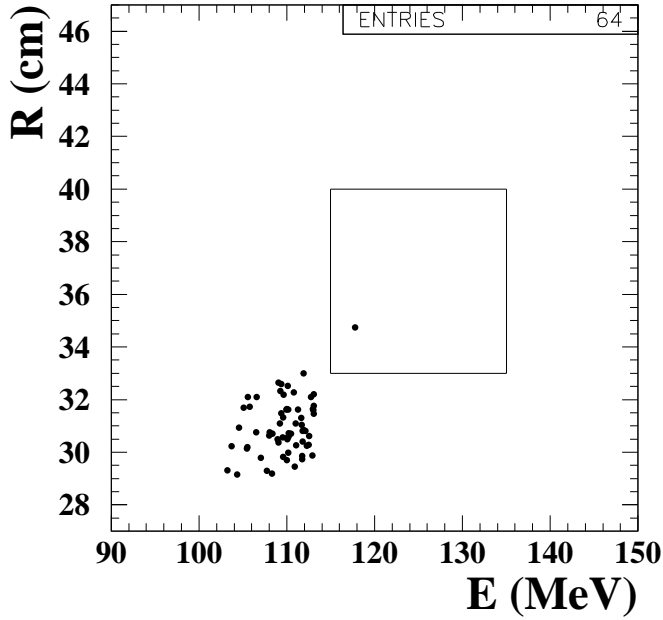}
    \caption {\it Result for 1995-97 datasets
    \label{newdata} }
\end{figure}

In Fig.\ref{newdata}, we present the results from the 1995-97
datasets. The plot shows the data after all cuts, except
those on Energy and Range of the $\pi^+$, have been made. We have one
event in the signal region. Using the total luminosity and detection
efficiency, we obtain\cite{second}

\begin{equation}
{\cal B}(K^+ \rightarrow \pi^+ \nu {\overline \nu}) = 1.5^{+3.4}_{-1.2} \times 10^{-10}
\label{result} 
\end{equation}

Assuming unitarity of the CKM matrix and $V_{cb}$, this result
implies, \\ $0.002 \le|$\vtd $|\le 0.04$.

\section{Results on Radiative Decays}

\subsection{\kppg}

  We have 19,836 events in this decay channel\cite{kppg}, which
represents an increase of a factor of eight in statistics over the
previous best measurement.

\begin{figure}[ht]
 \begin{minipage}{0.3\linewidth}
  \includegraphics{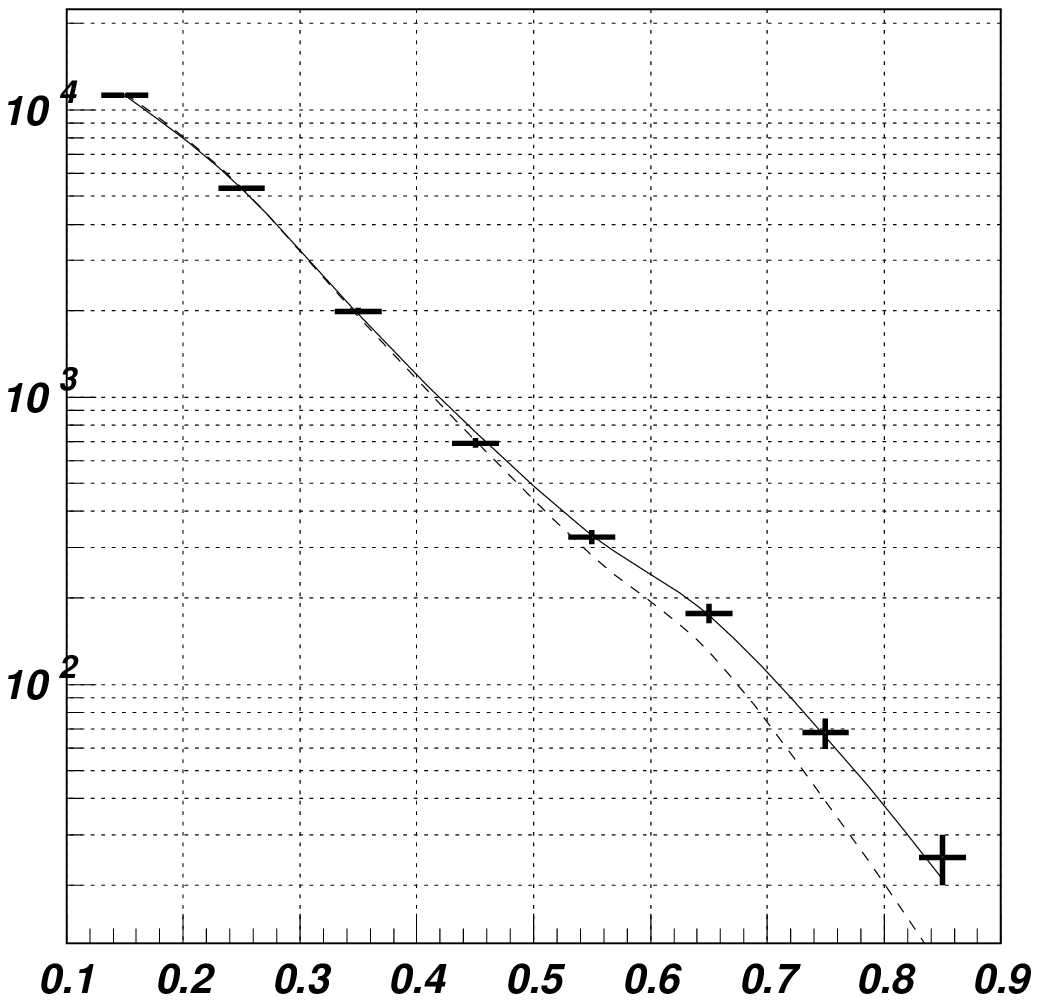}
 \end{minipage}\hfill
 \begin{minipage}{0.3\linewidth}
  \includegraphics{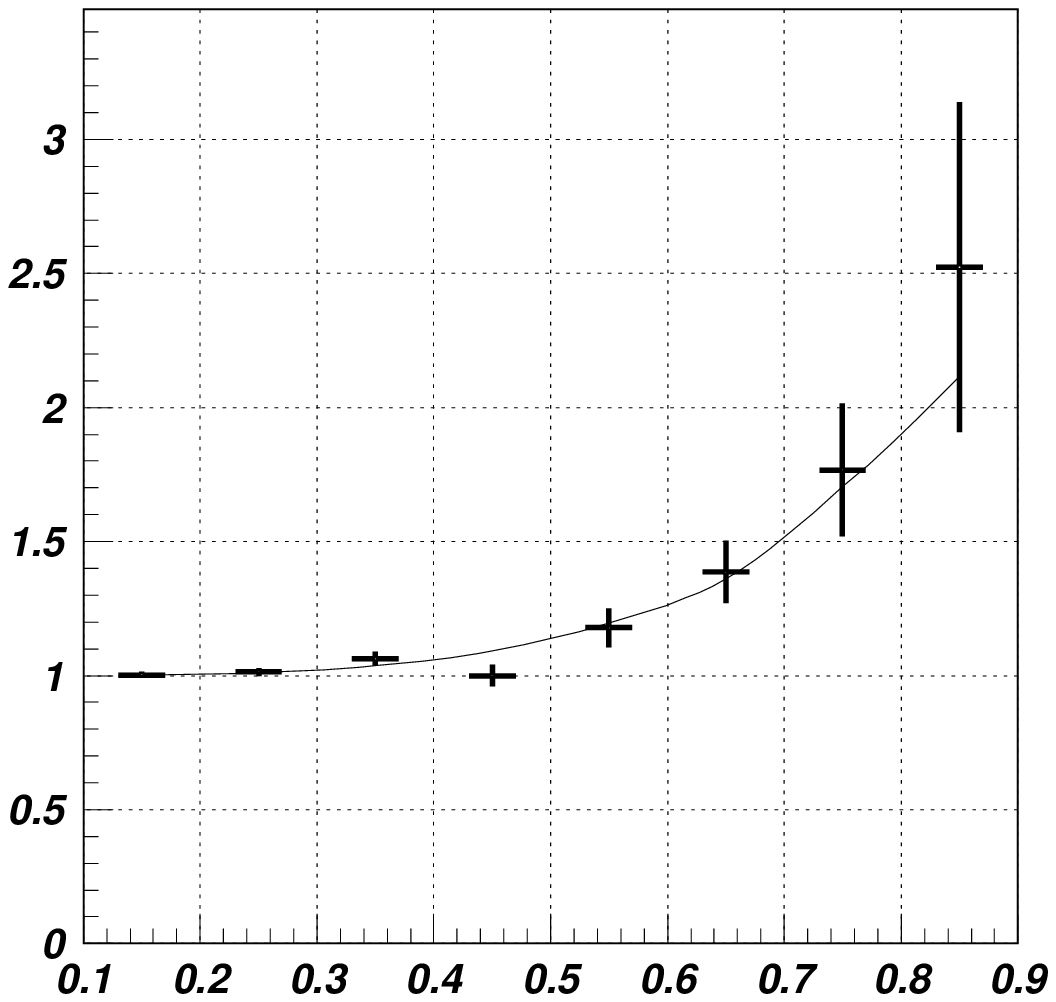}
 \end{minipage}\hfill
 \caption{\it (left) W spectrum of the signal events and best fits to
 IB+DE [solid curve] and IB alone [dashed curve]; (right) W spectrum
 normalized to the IB spectrum.  \label{fig_kppg} }
\end{figure}

The data is expressed in terms of a variable W, which is defined as, 

\begin{eqnarray}
{\rm W^{2}} & \equiv & {\rm (p\cdot q)/{m_{K^+}^{2}} \times
(p_{+}\cdot q)/{m_{\pi^+}^{2}}} \\ \nonumber
           &   =  & {\rm E_{\gamma}^2\times (E_{\pi^+} - P_{\pi^+}\times
\cos{\bf\theta_{\pi^+\;\gamma}})/({m_{K^+}^2}\times{m_{\pi^+}^{2}})}
\label{eqn_w}
\end{eqnarray}

\noindent where, p, p$_+$ and q are 4-momenta for the \kplus, $\pi^+$ and gamma,
respectively.  The structure dependent (DE) contribution is mainly at
high values of W, as shown in Fig.~\ref{fig_kppg}. Low values of W are
populated by Inner Bremsstrahlung (IB) decays.

\vspace{3.5cm}

We measure, 
\begin{equation}
{\rm BR(K^+ \! \rightarrow \! \pi^+ \pi^0 \gamma;DE)=
(4.72\pm0.77)\times 10^{-6} \, (55<\!T_{\pi^+}\!<90 MeV) }
\label{eqn_kppg}
\end{equation}

This result is roughly a factor of 4 lower than previous measurements.
In addition, the interference, between IB and DE, is measured to be
$(-0.4\pm 1.6)\%$, and the ratio, DE/IB is measured to be $(1.85\pm
0.30)\%$.

The decay rate, corrected to full phase space\footnote{This correction
assumes that the form factor has no energy dependence.}, is now
measured to be similar to that for $K_L$: $\Gamma(K^+ \! \rightarrow
\! \pi^+ \pi^0 \gamma;DE)= (808\pm132) s^{-1}$ vs. \\ $\Gamma(K^0_L
\! \rightarrow \! \pi^+ \pi^- \gamma;DE)= (617\pm18) s^{-1}$.

\subsection{\kmng}

The structure dependent (SD) contribution has two components, due to
the two polarizations of the outgoing photon. SD$^+$ peaks at higher
$E_\mu$ and E$_\gamma$ and is easier to detect. We have roughly 2700
events\cite{kmng} in the region of interest, E$_\gamma \ge 90 MeV$ and
$E_\mu \ge 137 MeV$.

\begin{figure}[ht]
\includegraphics{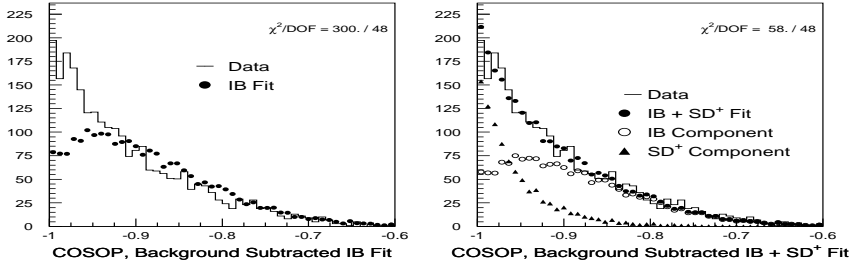}
\caption{\it Distribution of $\cos \theta_{\mu \gamma}$ for E787 \kmng\
candidates. A fit to IB alone is shown in (a), and a fit to both IB
and SD$^+$ is shown in (b).}\label{e787_kmng}
\end{figure}

In Figure~\ref{e787_kmng}, we present the data in terms of the angle
between the $\mu^+$ and the $\gamma$. It is clear from the fits that
there is a large contribution from SD.

\vspace {3.5cm}

The final result is based on a fit which includes contributions from
SD$^-$, IB and interference terms,

\begin{equation}
B(SD^+)  =  (1.33\pm0.12\pm0.18)\times10^{-5} 
\end{equation}

The sum of form factors is measured to be, $|F_V+F_A| =
0.165\pm0.007\pm0.011$.  This can be compared to the $O(p^4)$ $\chi$PT
calculation\cite{bijens} of $F_V+F_A = 0.137\pm0.006$ and
B(SD$^+$)=9.22$\times10^{-6}$.  We also measure $-0.04 \le F_V-F_A <
0.24$ at 90\% CL; this is to compared with an expectation\cite{bijens}
of ($0.052\pm 0.006$).

\section{Future Plans}

\subsection{1998 dataset}

As mentioned in Section~\ref{section:dataset}, we also took data in
1998. The sensitivity of this data set is expected to be approximately
the same as that of the combined 1995-97 datasets. This result is
expected in 2001.

\subsection{Phase Space below $K_{\pi2}$ peak}\label{section:kp2}

In the analysis presented here, we have searched for the signal in the
region above the $K_{\pi2}$ and below the $K_{\mu2}$ momentum
peaks. However, because of the V-A nature of the the decay, there is
also a large phase space below the $K_{\pi2}$ peak, as shown in
Fig.\ref{phase}. Some of the advantages of looking in this region are,
phase space is larger than above $K_{\pi2}$, less $\pi^+$ absorption
losses, etc., leading to a possibility of $\times 3$ increase in
acceptance.

However, the background due to $K_{\pi2}$ is very large, since the
$\pi^+$ can elastically scatter in a fiber hit by the incoming \kplus,
lose energy and fall inside the signal region.  Since this is a real
pion, we cannot use the kinematic discriminant to reject this
background. This scenario occurs when the daughter pions travel along
the beam axis. Since photon veto capability is weak in this angular
region, it is more likely to miss the two decay photons. One has to
look for evidence of the pion scatter in the kaon fiber, as a way to
discriminate against this background.

We made good progress in studying this region using the 1996
dataset. Searching in the momentum region, 140 MeV/c $ \le {\rm P}_\pi
\le$ 190 MeV/c, preliminary studies showed that S$_{\rm SM}$/B was
$\sim$ 1:5, where S$_{\rm SM}$ is the expected Standard Model signal.
We expect this ratio to improve to $\sim$ 1:3 when we extrapolate to
1995-98 datasets, and to $\sim$ 1:1 in experiment E949.

\subsection{Experiment E949}

    Experiment E949 is an upgrade of E787 and is scheduled to run in
2002-2003, with an engineering run scheduled for 2001. 

Many subsystems in E787 have been upgraded, e.g., {\it Photon veto}
has been improved by adding more detectors in the central region,
converting the current degrader into an active degrader, adding a
photon veto detector downstream of the target,etc., the {\it Range
Stack} has been improved by replacing some of the scintillator
layers, {\it Trigger} and {\it DAQ} systems have been upgraded,
etc. These improvements are expected to increase sensitivity by about
a factor of 3 (over the result based on the 1995 dataset\cite{first}).

In addition, the accelerator is scheduled to deliver more protons per
spill, increase the duty cycle and reduce the \kplus \space momentum,
leading to an improvement of a factor of 2.2 in sensitivity.

We plan to reoptimize our analysis for these higher rates and also
hope to gain sensitivity from looking in the phase space below the
$K_{\pi2}$ peak, as described in Section \ref{section:kp2}. Each of
these two sources is expected to yield an increase in sensitivity of a
factor of 2. However, we take a conservative approach and take only
one factor of 2 in our projections.

Factoring in the increase in running time ($\sim$ 6000 hours), we
estimate that our sensitivity will be a factor of 50 higher than the
result based on the 1995 dataset\cite{first}, which implies a Single
Event Sensitivity (SES) of about $8\times 10^{-12}$. This means that
we expect to observe 10 Standard Model events in E949, allowing us to
measure \vtd \space with a precision of about 23\%, which is extremely
competitive with the current precision of about 21\% obtained from
$B^0_{d} \overline B^0_{d}$ mixing.

\section{Conclusions}

  Using data collected in 1995-97, we have measured ,
\begin{equation}
{\cal B}(K^+ \rightarrow \pi^+ \nu {\overline \nu}) = 1.5^{+3.4}_{-1.2} \times 10^{-10}
\end{equation}

We have also made high statistics measurements of the structure
dependent parts of \kppg and \kmng.

Using all the data collected with the E787 detector, we expect to
improve the SES for \kpnn to $\sim 0.7\times 10^{-10}$.

Using the data which we expect to collect in experiment E949, we hope
to observe ${\cal O} (10)$ \kpnn Standard Model events. This will allow
us to measure \vtd \space with a precision of about 23\%.

\begin{figure}[ht] 
\vspace{7.0cm}
\includegraphics{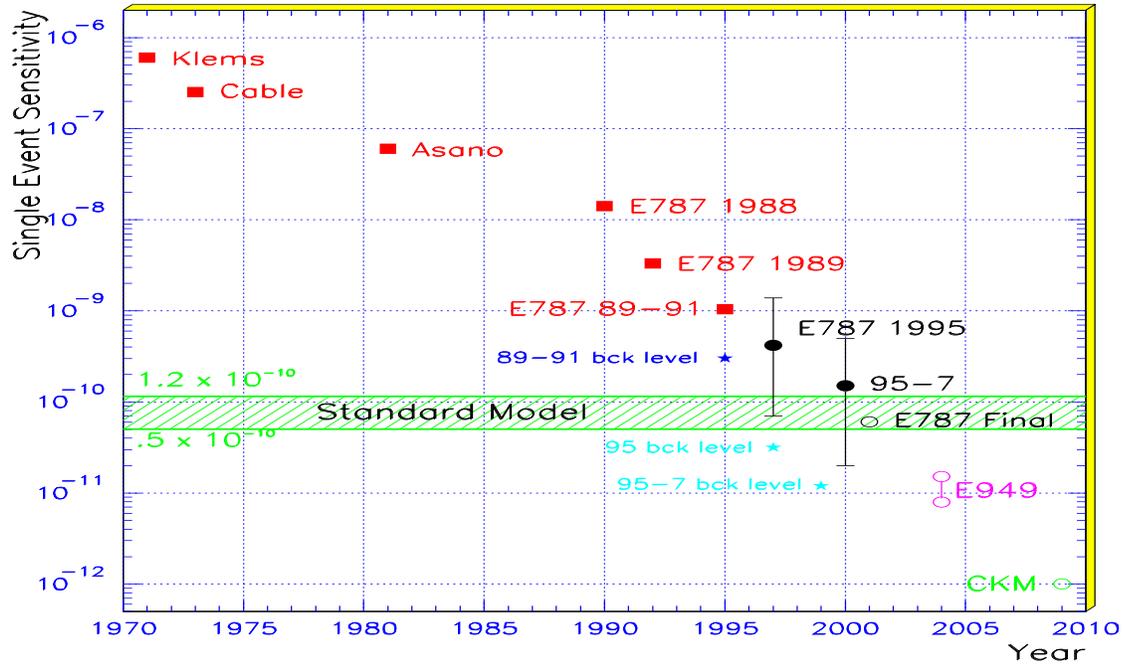}
    \caption {\it Single Event Sensitivity
    \label{SES} }
\end{figure}

In Fig.\ref{SES}, we present the improvement in SES. We also show the
expectations from the 1995-98 datasets (denoted as E787 final), from
experiment E949, and from a proposed experiment (CKM) at
Fermilab\cite{ckm}.

\newpage

\end{document}